\documentclass[]{aa}
\usepackage{psfig}
\usepackage{txfonts}
\begin{document}

\title{The dynamical evolution of fractal star clusters: the
survival of substructure}

\author{Simon\,P.\,Goodwin \and A.\,P.\,Whitworth}

\offprints{Simon.Goodwin@astro.cf.ac.uk}

\institute{Dept. of Physics \& Astronomy, Cardiff University, 5 The
Parade, Cardiff, CF24 3YB, UK}

\date{}

\abstract{

We simulate the dynamics of fractal star clusters, in order
to investigate the evolution of substructure in recently formed 
clusters. The velocity dispersion is found to be the key parameter 
determining  the survival of substructure. In clusters with a low 
initial velocity dispersion, the ensuing collapse of the cluster 
tends to erase substructure, although some substructure may 
persist beyond the collapse phase. In clusters with virial ratios 
of 0.5 or higher, initial density substructure survives for 
several crossing times, in virtually all cases. Even an initially 
homogeneous cluster can develop substructure, if it is born 
with coherent velocity dispersion.

These results suggest that the simple initial conditions 
used for many sophisticated $N$-body simulations could be 
missing a very important and dramatic phase of star cluster 
evolution.

}

\maketitle

\section{Introduction}

It appears that most stars -- possibly all stars -- form in
clusters. Their dynamical evolution is thus of great 
interest.  In recent years, codes such as {\sc nbody6} 
(Aarseth 2000), which include detailed stellar evolution and 
mass loss, binary evolution and mass transfer, have made 
possible a new generation of `kitchen sink' simulations (e.g. 
Kroupa et al. 2001; Hurley et al. 2001; Portegies
Zwart et al. 2001). However, the initial conditions for these 
simulations have often been very simple, for example Plummer 
models, in stark contrast with the great detail invoked in 
modelling the subsequent evolution.

Observations of star clusters, on the other hand, suggest that 
the initial conditions for star formation are highly clumpy and
structured, both in the distribution of the molecular gas from 
which  stars are about to form (e.g. Williams 1999 and references 
therein), and in the distribution of newly-formed stars (e.g. 
Bate et al. 1998; Gladwin et al. 1999).

Aarseth \& Hills (1972) were the first to investigate the 
evolution of collapsing star clusters with substructure. Their 
simulations were limited by the available computer power to 
clusters of 120 stars. They found that subclustering was destroyed 
on a free-fall timescale. Later Goodwin (1998) investigated 
an initially virialised cluster with density substructure and 
a larger number of stars. He found that most of the initial 
substructure was erased within a few crossing times.

In this paper we investigate the evolution of initially fractal 
star clusters to see how long substructure can survive.  Our 
models include star clusters with large velocity dispersions, 
as would be expected in clusters shortly after they expel their 
residual gas (cf. Goodwin 1997).  In addition, we investigate 
fractal clusters in which the density substructure is 
correlated with coherent velocity dispersion, as would be 
expected in clusters where sub-clusters form from distinct 
molecular cores.

We are not suggesting that star clusters {\em are} necessarily 
fractal. The range of scales over which young star clusters 
exhibit substructure is usually very small, often less than an 
order of magnitude, so the notion of a fractal cannot be applied 
rigorously. Nonetheless, fractals provide a simple, one-parameter 
description of clumpiness, and this is why we are using them.

\section{$N$-body method and initial conditions}

We conduct our $N$-body simulations on a {\sc grape-5} board, 
which allows for very rapid solution of the $N$-body 
gravitational problem (Kawai et al. 2000).  We use a simple 
direct first order $N$-body integrator, as the speed of the 
{\sc grape-5} board allows the timestep to be set sufficiently 
small that over the course of a simulations the total energy 
of the system never changes by more than 0.01 per cent, and 
usually by significantly less than this. A small softening 
length is used (generally of order $10^{-5}$ in code units). 

The presence of binaries in a cluster will effect the dynamical
evolution of the system by altering impact parameters.  However, 
tests using different softening lengths show the results to 
be independent of the softenning length, even when it is 
significantly larger than our canonical $10^{-5}$.  This is due to the
relaxation being primerally violent relaxation, rather than
encounter-driven two-body relaxation.

Fractal distributions are generated by defining an {\it ur}-cube 
with side 2, and placing an {\it ur}-parent at the centre of the 
ur-cube. Next, the ur-cube is divided into $N_{\rm div}^3$ equal 
sub-cubes, and a child is placed at the centre of each sub-cube 
(the first generation). Normally we use $N_{\rm div} = 2$, in 
which case there are 8 sub-cubes and 8 
first-generation children. The probability that a child matures 
to become a parent in its own right is $N_{\rm div}^{(D-3)}$, 
where $D$ is the fractal dimension; for lower $D$, 
the probability that a child matures to become a parent is lower. 
Children that do not mature are deleted, along with the ur-parent. 
A little noise is then added to the positions of the remaining 
children, to avoid an obviously gridded structure, and they 
become the parents of the next generation, each one 
spawning $N_{\rm div}^3$ children (the second generation) at the 
centres of $N_{\rm div}^3$ equal-volume sub-sub-cubes, and each 
second-generation child having a probability $N_{\rm div}^{(D-3)}$ 
of maturing to become a parent. This process is repeated recursively 
until there is a sufficiently large generation that, even after 
pruning to impose a spherical envelope of radius 1 within the 
ur-cube, there are more children than the required number 
of stars. Children are then culled randomly until the required 
number is left, and the survivng children are identified with 
the stars of the cluster.

We explore a range of models with $N_{\rm tot}=1000$ and 
$N_{\rm tot}=10000$ stars. The fractal dimensions which we 
investigate are $D = 1.6,\,2.0,\,2.6,\,\&\,3.0$, since these 
all correspond to $2^D$ (the mean number of maturing children) 
being close to an integer. This reduces the likelihood 
of departures from the specified fractal dimension, because 
in our algorithm for constructing an initial cluster  
the number of maturing children born to each parent should be 
an integer.  

Our simulations begin with a random velocity dispersion, which is 
either incoherent or coherent. For an incoherent velocity dispersion, 
each particle is given random cartesian velocity components from a 
gaussian distribution, and these velocities are then scaled so that 
the virial ratio $\alpha$ has the prescribed value. $\alpha$ is 
defined as the ratio of the total kinetic energy to the magnitude 
of the gravitational potential energy. $\,\alpha = 0.1$ corresponds 
to a cluster which immediately collapses. $\,\alpha = 0.5$ corresponds 
to virial equilibrium. $\,\alpha = 0.75$ corresponds to a 
super-virial cluster (i.e. one which has just expelled its residual 
gas).

For a coherent velocity dispersion, each star inherits most of 
its velocity from its family tree. We first calculate, for each 
mature child of each generation, how many stars are descended 
from it, and this gives the child's mass. Next, starting with 
the children of the first generation, we give them random 
velocities relative to the ur-parent, so that they form a 
virialised cluster. We treat the children of each subsequent 
family in the same way, giving them random velocities relative 
to the parent, so that they form a virialised sub-cluster, but 
in addition they acquire the velocity of the parent. This 
process is repeated recursively, and hence the stars of the 
final generation inherit  random velocities from all their 
antecedents, giving a coherent velocity dispersion. Finally, 
the velocities are scaled so that the virial ratio $\alpha$ 
has the prescribed value.

We believe that the most realistic of the initial velocity 
dispersions we use is the coherent, super-virial one 
($\alpha = 0.75$). Our simulations do not include a gaseous 
component, but we assume that the residual gas has been 
expelled from the cluster very rapidly and recently. If 
the stars have previously been in virial equilibrium with 
the residual gas, the virial ratio of the purely stellar 
component is super-virial. It also seems likely that the 
velocities of the stars will be correlated locally, on the 
assumption that each sub-cluster of stars has formed from 
a single molecular core.

\section{Measures of clumpiness}

The primary aim of this paper is to investigate how rapidly 
initial density substructure and associated velocity coherence 
are destroyed, and the typical timescale before a cluster can 
be described as smooth. In order to do this we require a 
robust method for measuring the clumpiness of a cluster.

One method would be to determine the evolution of the fractal 
dimension of the cluster. However, this method has drawbacks, 
because it is difficult to measure the fractal dimension of a 
distribution with a strong overall radial density gradient. 
Clusters rapidly acquire a core-halo structure, in which the 
mean separation between stars decreases radially by a 
significant factor. To determine the fractal box-dimension of 
a cluster, a grid of cells is overlaid on the cluster, and 
the number of cells containing at least one star is counted. 
This is repeated, starting with a coarse grid of large cells, 
and proceeding to ever finer grids of smaller cells. The 
fractal box-dimension is then given by the slope of a plot of 
the log of the number of occupied cells against the log of 
the inverse cell size. This slope turns over at the point 
where the grid becomes saturated, i.e. where the number of 
occupied cells is equal to the number of stars; further 
reductions of cell size then make no difference to cell 
occupancy. However, in the presence of a density gradient,  
saturation of the halo occurs before saturation of the core, 
producing a complicated relationship that is difficult to 
interpret. It might be possible to compensate for this, if 
the overall radial density distribution is known a priori, 
but in general it is not.

The simple method we use to measure clumpiness is to accord 
each star a local density, given by $5m/V_5$, where $m$ is 
the stellar mass and $V_5$ is the spherical volume bounded 
by the fifth nearest star. Next we overlay spherical shells 
centred on the centre of mass of the ten densest stars, and 
work out the average density in each shell. Substructure is 
then visible as stars (or groups of stars) within a shell 
having density significantly higher than the average density 
for that shell. As a measure of the level of clumpiness, we 
use two numbers, $F_{20}$ and $F_{50}$. $F_{20}$ ($F_{50}$) 
measures the fraction of shells in which more than 20\% 
(50\%) of the stars have more than 5 times the average 
density. A high level of clumpiness is reflected in large 
values of $F_{20}$ and $F_{50}$. We tested more complicated, 
kernel weighted density estimates (such as those used in SPH 
simulations), but found no significant difference in the 
results.

\begin{figure*}
\centerline{\psfig{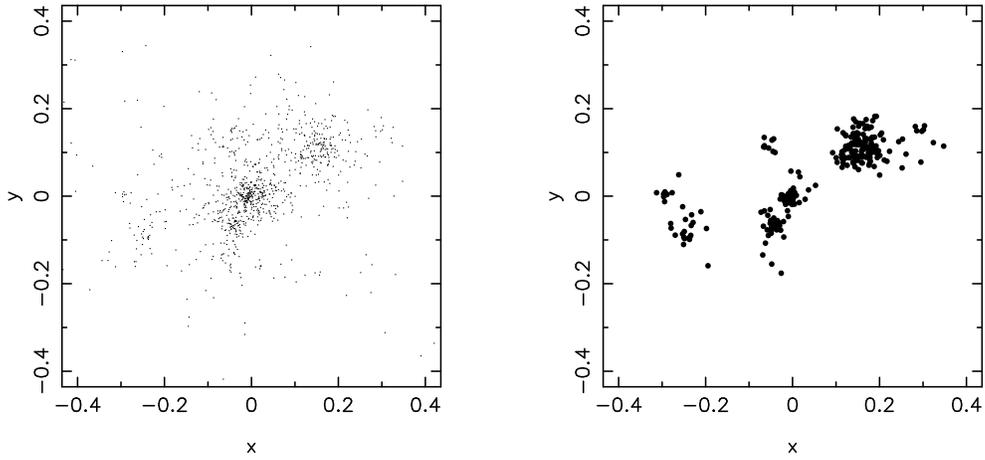}}
\caption{Left: a clumpy distribution of stars.  Right: the stars
selected by our clumpiness
measure as being significantly overdense.  This distribution has high
$F_{20}$ and $F_{50}$ which indicate that it is significantly clumpy
(which the eye confirms).}
\label{fig:clumpy}
\end{figure*}

$F_{20}$ and $F_{50}$ are not greatly affected by small 
number statistics. Tests performed with clusters of 1000 
stars show that $F_{50} < F_{20} \la 0.1$ for various 
clusters which should not have significant clumpiness, 
viz. a randomly distributed uniform-density cluster, a 
randomly distributed Plummer cluster, and a $D = 3$ 
fractal cluster. In contrast, a $D = 1.6$ fractal 
cluster, which should be very clumpy, has $F_{50} 
\sim F_{20} \sim 1$. We conclude that $F_{20}$ and 
$F_{50}$ are useful measures of clumpiness. 
Figure~\ref{fig:clumpy} shows a fractal distribution with 
obvious substructure, and the large points show which stars 
have been selected by this method as being 5 times the 
average density. They coincide well with the substructure 
which the human eye identifies.

\section{The evolution of fractal clusters}

We now use $F_{20}$ and $F_{50}$ to investigate the evolution 
and erasure of substructure from an initially fractal cluster.

\subsection{Collapsing clusters}

\begin{figure*}
\centerline{\psfig{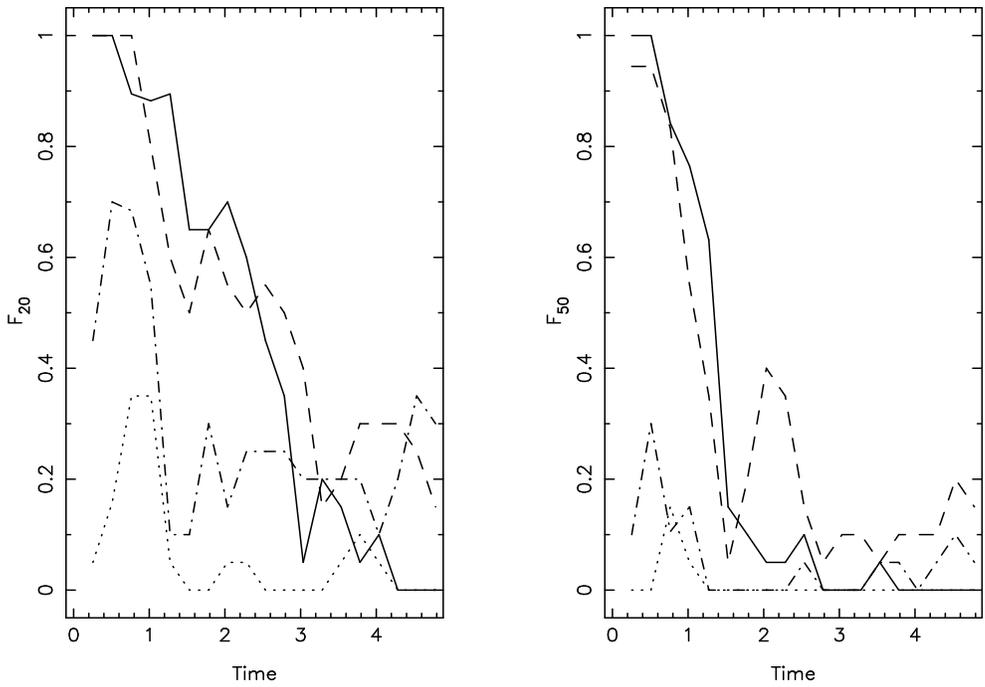}}
\caption{The evolution of the $F_{20}$ and $F_{50}$ measures of
clumpiness for clusters with 1000 particles, initial virial ratio 0.1,
coherent velocities and initial $D$ of 1.6 (solid lines), 2
(dashed lines), 2.6 (dot-dashed lines) and 3 (dotted lines).}
\label{fig:erase}
\end{figure*}

Figure~\ref{fig:erase} shows typical results for the evolution 
of $F_{20}$ and $F_{50}$ for clusters having initial virial 
ratio $\alpha = 
0.1$ (such that they will collapse) and initial fractal 
dimensions $D = 1.6,\,2.0,\,2.6,\,\&\,3$ (decreasing 
levels of initial substructure).  In all of these cases the 
initial velocity dispersion is coherent.

Time is given in $N$-body units, such that $G=M=R=1$, where 
$M$ and $R$ are the total mass and initial radius of the 
cluster. For example, if the total mass of the cluster is 
$M = 100 M_{\odot}$ and the initial radius is 
$R = 1\,{\rm pc}$, then one time unit is $\approx 1.5$ Myr 
(see Heggie \& Mathieu 1986 for details of $N$-body units).

Figure~\ref{fig:erase} shows that the level of substructure 
tends to decrease with time, and all substructure has 
essentially disappeared by $T=3$. An interesting feature 
is the transient rise in the level of substructure in 
clusters with high $D$ (i.e. those with initially low
levels of substructure). This is due to the coherence in 
the initial velocity field. Even though there is initially 
a low level of density substructure, for a short time the 
coherent velocity field generates substructure. However, 
this substructure does not last long.

The main mechanism for erasing substructure is the 
gravitational interactions between clumps.  The potential 
of a clumpy cluster is highly uneven and violent relaxation 
occurs, allowing the stars to relax into a smooth distribution 
on a short timescale. Two-body encounters also act to remove 
kinetic energy from the main cluster by ejecting stars. This 
can be seen in a rapidly expanding halo of stars around the 
main cluster, and a cluster core which is often 
significantly smaller than the initial size of the system.

\begin{figure*}
\centerline{\psfig{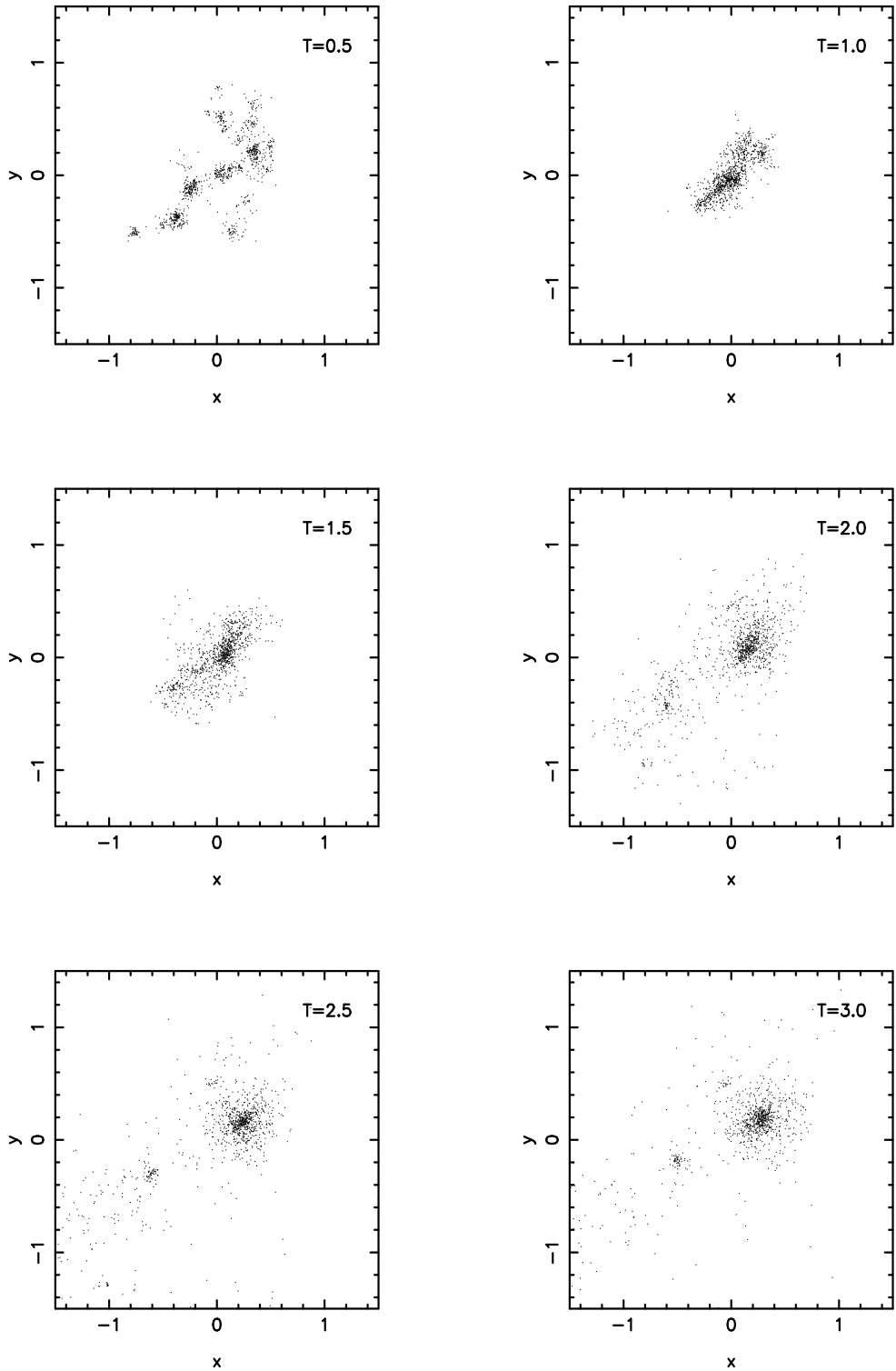}}
\caption{The evolution of an initially $F_{\rm dim} = 2$ cluster with
$N=1000$ and a virial ratio $\alpha=0.1$ with coherent velocities.  The time
in $N$-body units is given in the top right of each panel.}
\label{fig:evol}
\end{figure*}

Figure~\ref{fig:evol} shows in detail the evolution of the 
$D = 2$ cluster from the previous figure.  By inspection 
the evolution of the substructure is well described by our 
measures $F_{20}$ and $F_{50}$. In Fig.~\ref{fig:evol} the 
initially very clumpy distribution rapidly collapses. At 
first some of the clumps disperse, but those which are 
bound collapse and the larger of these clumps then attract 
nearby clumps. The initial clumpiness is erased 
as most clumps merge. However, one of the 
merging clumps is sufficiently bound to survive the initial 
merger process and emerges at late times to the bottom left.  
The re-emergence of this clump explains the increase in 
$F_{20}$ and $F_{50}$ for this cluster in 
Fig.~\ref{fig:erase} around $T=2$.

\subsection{Virialised clusters}

\begin{figure*}
\centerline{\psfig{figure=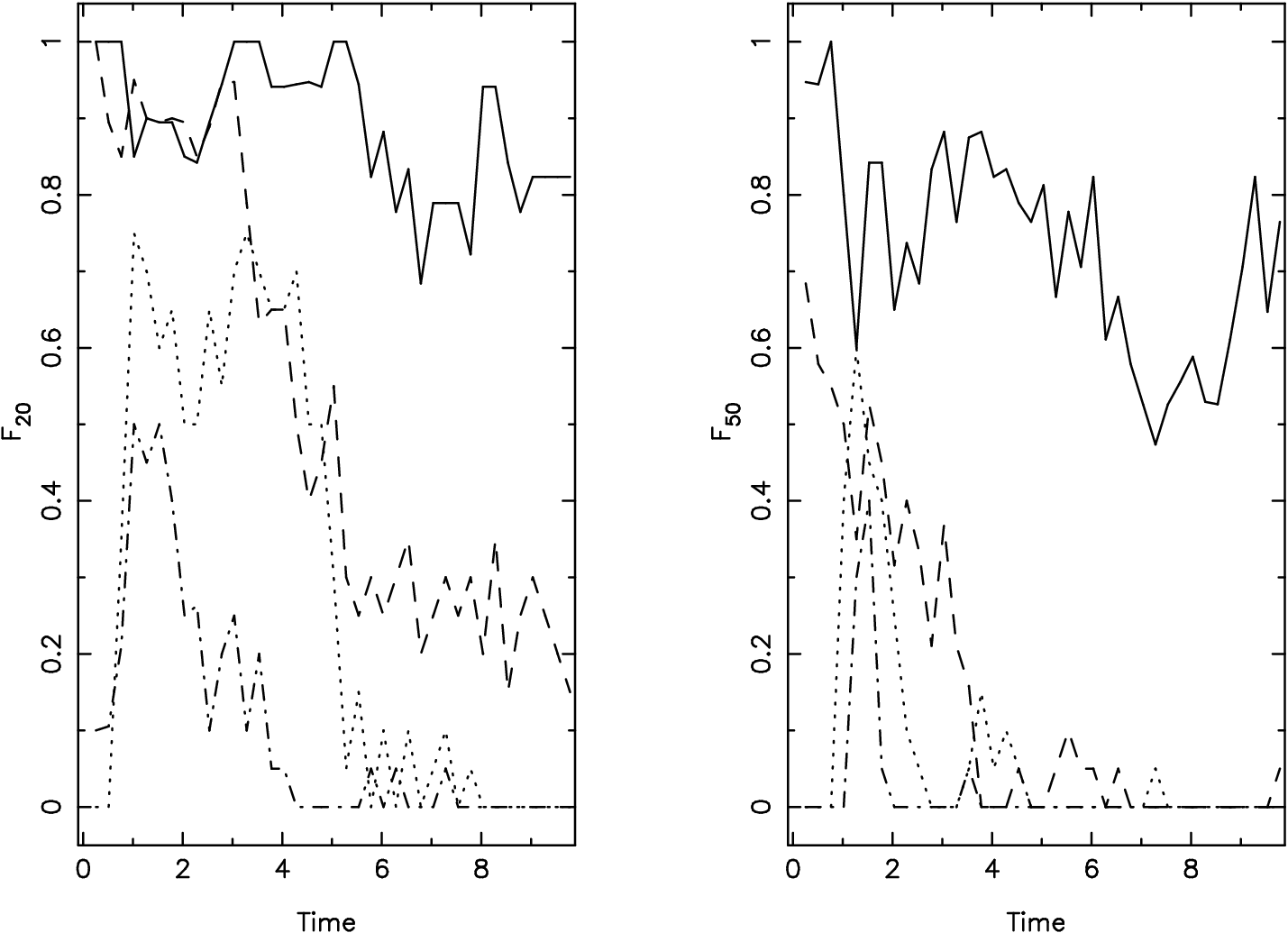,height=9.0cm,width=13.0cm,angle=270}}
\caption{The same as Fig.~\ref{fig:erase} but for clusters with an
initial virial ratio of $\alpha=0.5$.}
\label{fig:virial}
\end{figure*}

Figure~\ref{fig:virial} shows the evolution of $F_{20}$ and 
$F_{50}$, as in Fig.~\ref{fig:erase}, but for clusters with 
an initial virial ratio of $\alpha = 0.5$. Clusters with 
$\alpha = 0.5$ have enough kinetic energy to support 
themselves against overall collapse, but their clumpy
nature means that significant dynamical evolution still 
occurs. The most significant features of Fig.~\ref{fig:virial} 
are the longer time required for substructure to be erased, 
and the persistence of substructure throughout the simulation 
in the $D = 1.6$ cluster.

The initial coherence of the velocity dispersion in the 
high-$D$ simulations is again responsible for the increase 
in substructure early on, although this substructure is 
erased over a timescale of a few time units (a few Myr in a 
typical cluster).

\begin{figure*}
\centerline{\psfig{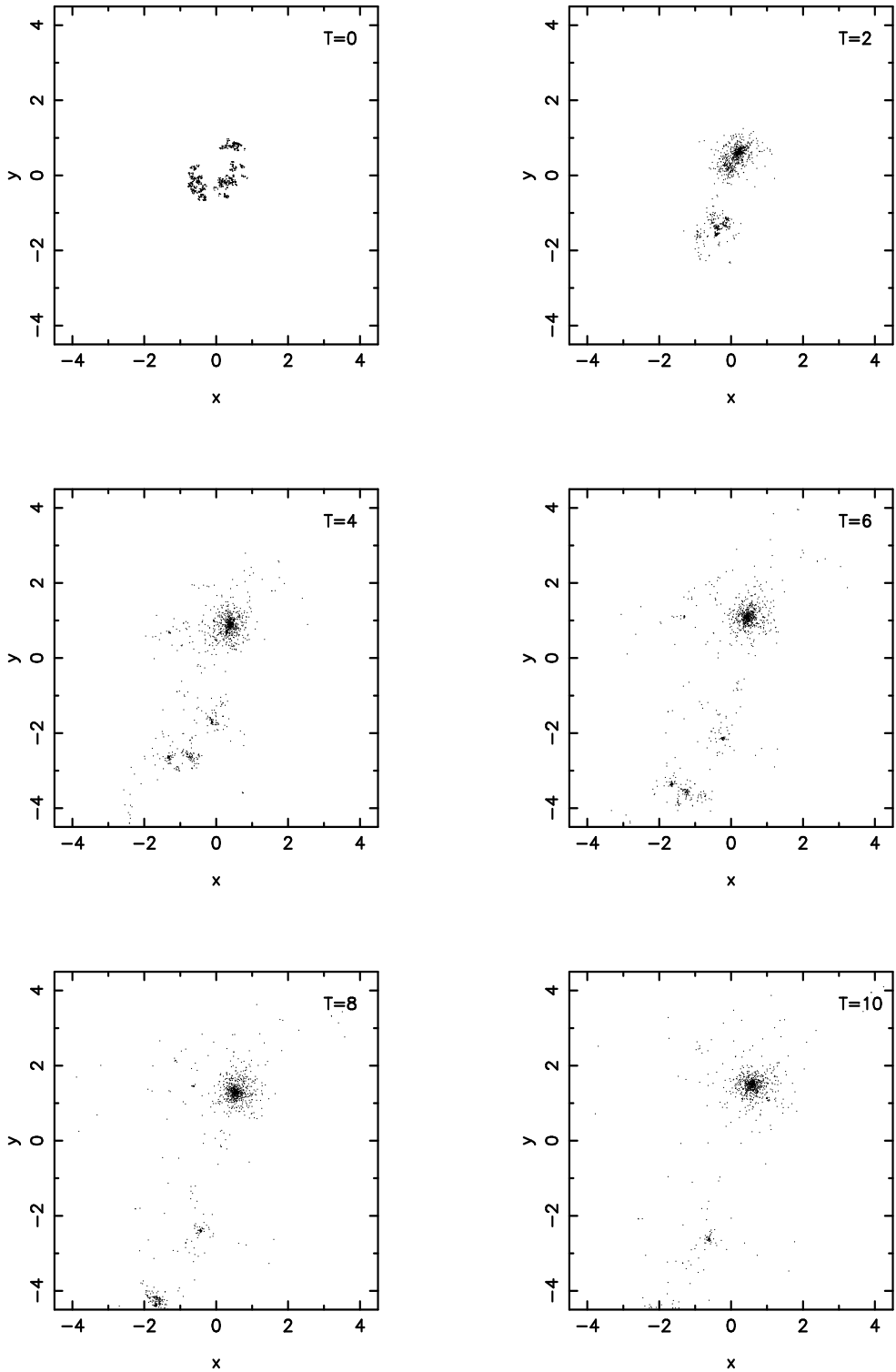}}
\caption{The same as Fig.~\ref{fig:evol} but for a cluster with an
initial virial ratio of $\alpha=0.5$.}
\label{fig:virevol}
\end{figure*}

In the $D = 1.6$ cluster the survival of substructure is due
to the high velocity dispersion and its coherence.  As can 
be seen in Fig.~\ref{fig:virevol} the initial cluster rapidly 
divides into a main cluster and a lumpy sub-cluster, which has 
sufficient bulk velocity to escape from the main sub-cluster, 
dividing into 3 small sub-sub-clusters as it does so.

Such an evolution is typical for clusters of low fractal 
dimension and not unusual for clusters of higher fractal 
dimension.  The key is the coherent nature of the velocity 
dispersion. The velocity dispersion {\it within} a 
sub-cluster holds it up against collapse, and its bulk 
velocity helps it to avoid merging with other sub-clusters. 
Hence it is able to maintain its separate identity for a 
long time. This contrasts with the cases discussed in the 
previous subsection, where the collapse tends to erase 
substructure rather quickly.

\subsection{Supervirial clusters}

\begin{figure*}
\centerline{\psfig{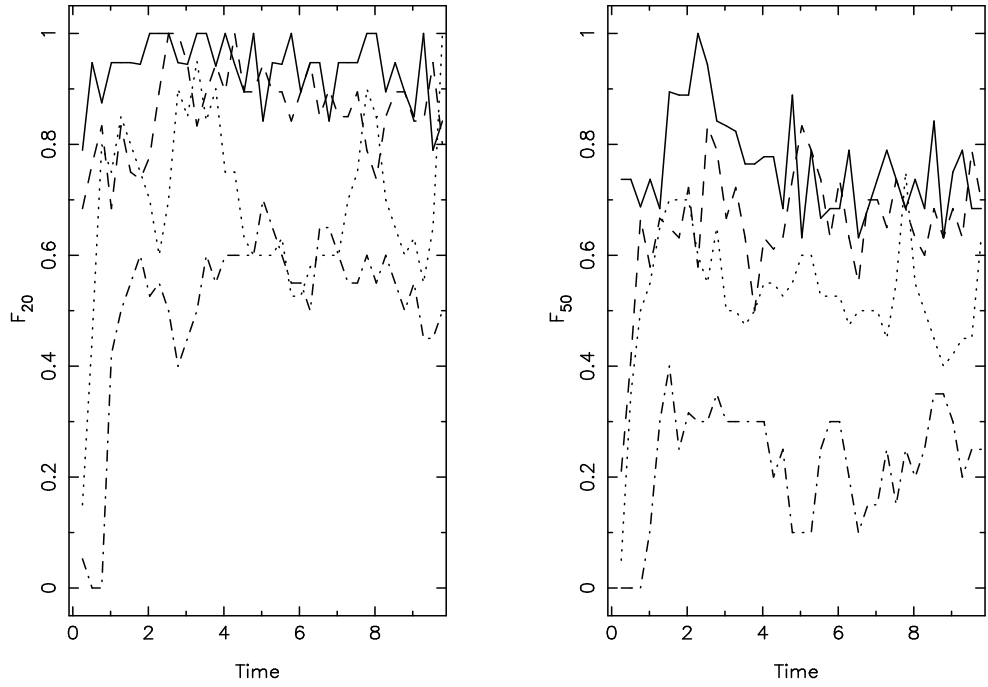}}
\caption{As Fig.~\ref{fig:erase} but for clusters with an
initial virial ratio of $\alpha=0.75$.}
\label{fig:super}
\end{figure*}

Figure.~\ref{fig:super} again shows $F_{20}$ and $F_{50}$, this 
time for clusters with an initial virial ratio of $\alpha = 0.75$, 
such that they expand.  In these cases it is clear that the level 
of substructure does not decrease rapidly. Significant substructure 
remains at the end of the simulations for all $D$.

\begin{figure*}
\centerline{\psfig{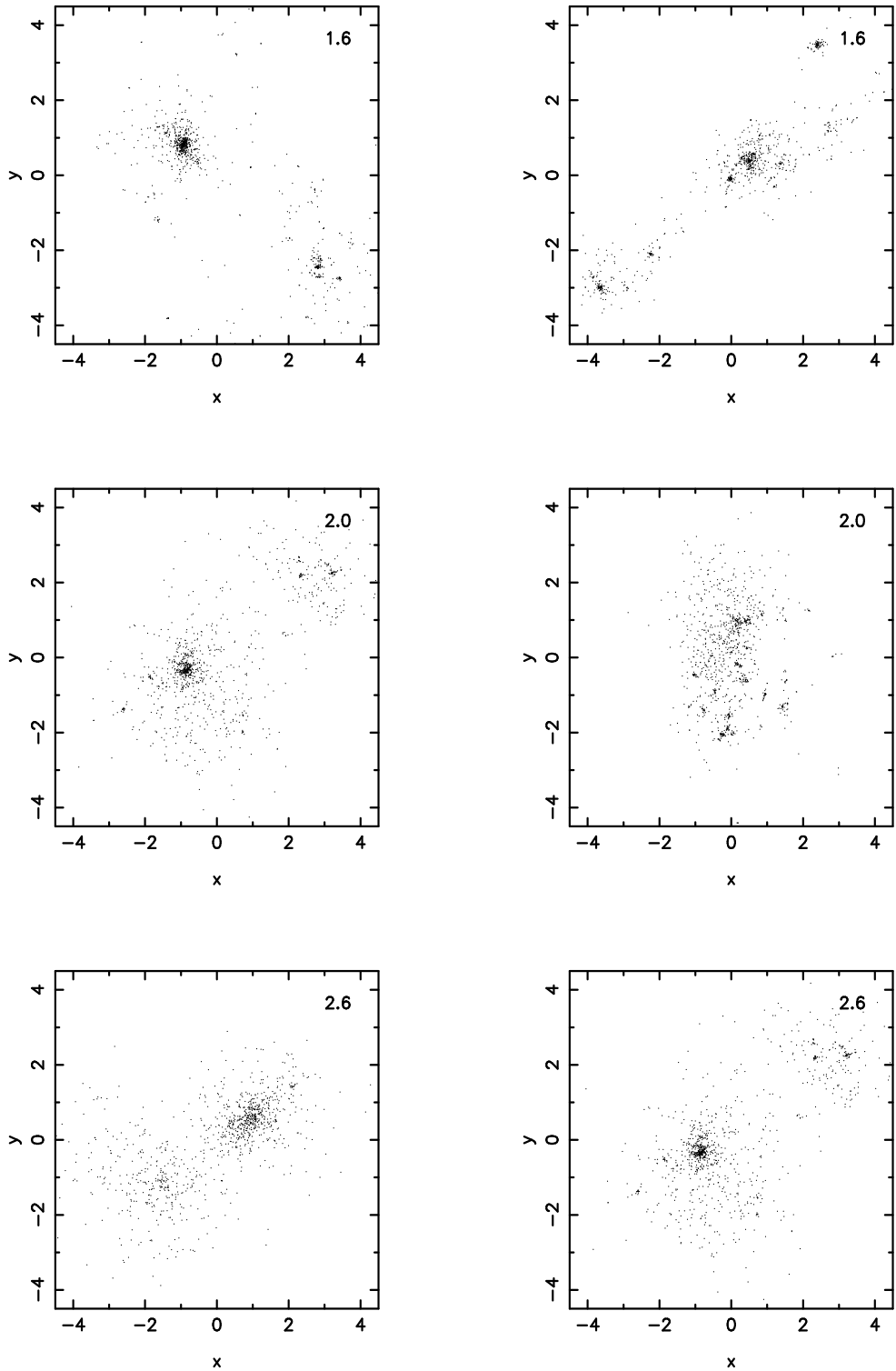}}
\caption{The final states of different realisations of clusters 
at $T=5$, with $N=1000$, virial
ratios of $\alpha=0.75$ and coherent velocities.  The initial fractal 
dimension is labeled in each case in the top right.}
\label{fig:final}
\end{figure*}

Even clusters whose density structure is initially not very 
clumpy grow density substructure, again due to the coherence of 
their initial velocity dispersion. Generally, the lower $D$, the 
more clumps there are, and the smaller they are. For high $D$ 
often a binary cluster is formed with two major sub-clusters. 
Figure~\ref{fig:final} shows the final states (at $T=5$) for 
different realisations of clusters with a variety of initial fractal 
dimensions.  In virtually all simulations clusters still have a 
significant amount of well defined substructure present at $T=5$. 
(For total mass $M \approx 1000 M_\odot$ and initial radius 
$R \approx 1\,{\rm pc}$, $T = 5$ corresponds to 5 to 10 Myr.)

Typically, by $T=5$ a main sub-cluster is identifiable (as a 
sub-cluster that is significantly larger than all other 
sub-clusters). Figure~\ref{fig:final} shows in most cases a 
clear main sub-cluster with surrounding substructure.  In many 
cases the substructure is never erased altogether; some clumps 
are not bound to the main clump and so escape.  This is very 
common in clusters with an initial virial ratio of $\alpha = 
0.75$, and even occurs occasionally in clusters with an 
initial virial ratio of $\alpha = 0.5$. Sub-clusters 
that {\it are} bound to the main sub-cluster can remain 
as separate entities for a significant length of time. 
We ran a subset of our simulations until $T=100$ and found some 
sub-clusters remaining in orbit. However, in the majority of 
cases the tidal effect of the main sub-cluster disrupts any bound 
substructure by $T=10$ or $20$ (roughly $10$ to $50$ Myr). 
This leaves a smooth, but typically elliptical, main cluster 
surrounded by a large, and expanding, halo of stars.  We did 
not however investigate the details of this phase of cluster 
evolution in any detail.

\subsection{Incoherent velocity structures}

\begin{figure*}
\centerline{\psfig{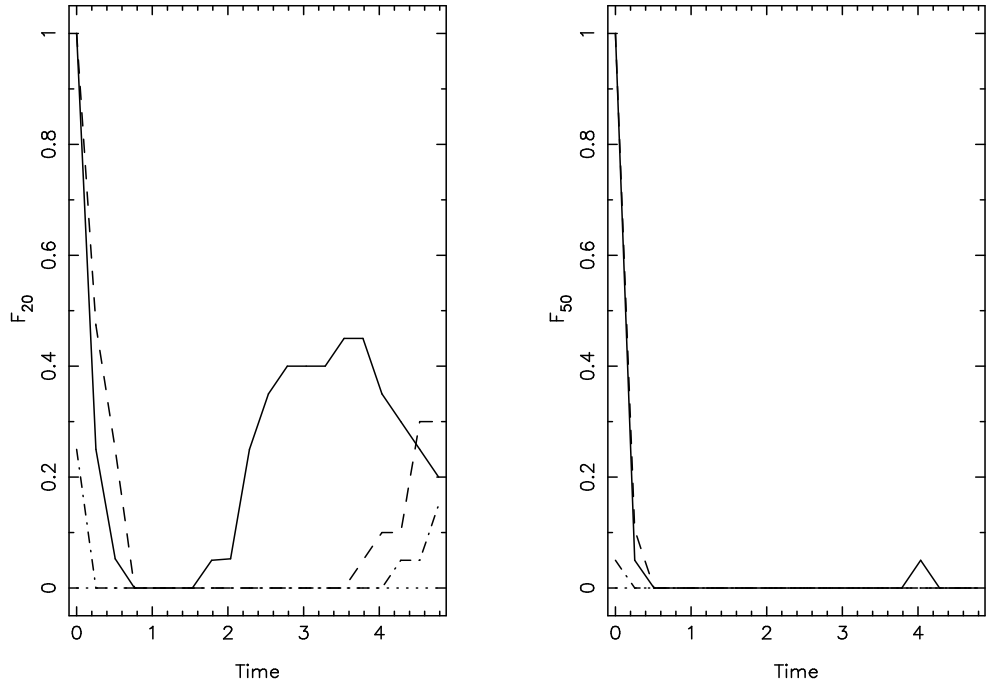}}
\caption{As Fig.~\ref{fig:erase} but for clusters with an
initial virial ratio of $\alpha=0.75$ and a random rather than coherent
velocity substructure.}
\label{fig:random}
\end{figure*}

Given the potential importance of coherent velocity 
dispersion in maintaining -- and even increasing -- the 
level of substructure, it is expected that clusters with 
incoherent velocity dispersion will not retain substructure 
so long. Figure~\ref{fig:random} shows the evolution of 
$F_{20}$ and $F_{50}$ for a cluster with initial virial 
ratio $\alpha = 0.75$ and incoherent velocity dispersion. 
The substructure is erased almost immediately. This is 
unsurprising, as, with such a large virial ratio,
any density substructure almost instantly disperses.

However, we believe that coherent velocity dispersion 
is the more plausible initial condition. The likely 
cause of the substructure that is commonly observed in 
young star clusters is the formation of sub-clusters of 
stars in distinct molecular cores formed by the turbulence 
in the molecular cloud. It is therefore to be 
expected that the stars within a sub-cluster have similar 
velocities, and that the velocity dispersion is correlated 
with the density substructure.

\subsection{The effect of large $N$}

The previous results have concentrated on clusters with $N_{\rm tot}=1000$ 
representing moderately rich, open clusters (for example, 
the Pleiades or Orion). Here we investigate the effect of larger 
$N_{\rm tot}$ on the survival or destruction of substructure.

\begin{figure*}
\centerline{\psfig{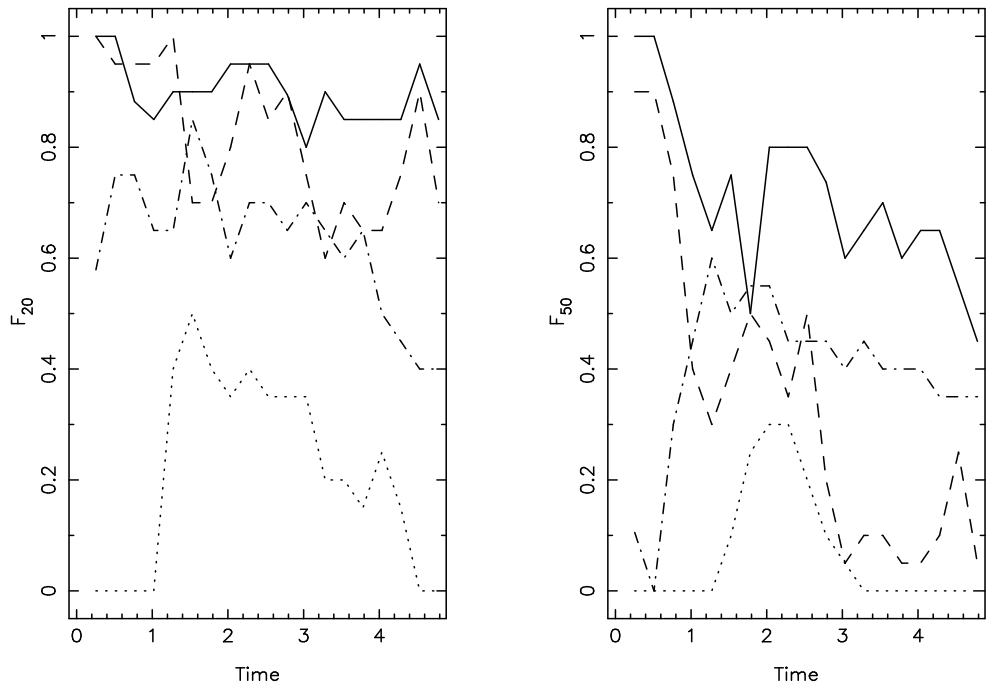}}
\caption{As Fig.~\ref{fig:super} but for clusters with $N=10000$.}
\label{fig:bign}
\end{figure*}

Figure~\ref{fig:bign} shows the evolution of $F_{20}$ and 
$F_{50}$ for clusters with virial ratio $\alpha = 0.75$, 
coherent velocity dispersion, and $D = 1.6,\,2,\,2.6,\,\&\,3$, 
equivalent to Fig.~\ref{fig:super} but with $N_{\rm tot}=10000$. 
Figure~\ref{fig:bign} illustrates a common theme for large
$N_{\rm tot}$, namely that the decline in $F_{20}$ and 
$F_{50}$ is somewhat more rapid than in clusters with 
small $N_{\rm tot}$.

This result is largely an artifact of the clump detection
proceedure.  Often in $N=1000$ clusters the number of particles within 
a particular radial bin is small (of order a few), whilst in $N=10000$
clusters that number is (obviously) far larger.  Thus poisson noise
has more of an effect upon the statistics in low-$N$ clusters.  

As an example, two $\alpha = 0.75$, $D = 2$ clusters were simulated, one
with $N=10000$ and the other with 90 per cent of the particles from
the first simulation removed at random.  At $T=5$, the $N=1000$ 
cluster had $F_{20}=0.85$ and $F_{50}=0.25$, and for
$N=10000$, $F_{20}=0.70$ and $F_{50}=0.05$.  The $F$-statistics in the
$N=1000$ simulation were increased by the effect of low numbers of
particles (5 to 10) in the inner radial bins, while the $N=10000$
cluster suffered far less from this effect with no less than 20
particles in any one bin.  The density of particles in low-$N$
simulations is also less smooth than in high-$N$ simulations as the
search radius in which 5 particles are to be found is usually
significantly greater.  Small clumps in low-$N$ simulations tend
to be more above the average density than in large-$N$ simulations. 

Nonetheless, large $N_{\rm tot}$ clusters still maintain 
significant levels of substructure for several crossing times. 
This may be an explanation for some of the anomalous bumps 
observed in the profiles of several LMC star clusters (e.g. 
Mackey \& Gilmore 2003).  

\subsection{The final structure of clusters}

An examination of the most significant sub-clusters at the end of
simulations shows that they generally appear similar to older
clusters.  The relaxation of the clusters creates Plummer- or
King-like profiles, ie. a flat central density profile, with a steep
decline in the halo.  This is especially clear in clusters with $N =
10000$.  The cluster illustrated in Fig.~\ref{fig:lostp}, has a flat
central surface density of $\approx 500$ stars per unit area over a
radius of $\approx 0.5$, followed by an approximate $r^{-2}$ decline.
There are significant sub-clusters apparent in the halo, the
largest of which cause `bumps' to appear in the surface density 
profile.  For clusters with low $N$, the trend to form core-halo
density structures is present, but far less clear due to 
low-$N$ noise.

Some clusters have distinct, unbound sub-clusters which are included
in our clumpiness determinations.  Once a sub-cluster has travelled a
significant distance from the main cluster it will be difficult
(without proper motions) to determine that it formed in the same
position as the main cluster.  Thus it may appear as if they are 
smooth and separate clusters that formed coevally, rather than the
result of dynamical segregation from the same initial cluster.

\section{Implications for observations and simulations}

Observations of young star clusters often show a very 
inhomogeneous, clumpy distribution.  Our simulations 
demonstrate that if the velocity dispersion of these 
clusters is low, then much of that initial substructure 
will be erased in the ensuing collapse (cf. Aarseth \&
Hills 1972). However, if, as we might expect, the velocity 
dispersion is high, such that the cluster remains 
supported or even expands, then significant levels of 
substructure can survive for several crossing times. Even 
an initially homogeneous distribution of stars can grow 
substructure if the initial velocity dispersion is coherent.

These results have two important consequences. First, 
young clusters probably undergo significant and rapid 
dynamical evolution. Therefore drawing conclusions about 
the initial stellar distribution from observations is 
very difficult, and should take these considerations into
account. Taurus is a good example of a young, embedded cluster with 
a high level of substructure (Briceno et al. 1993; Ghez 1993). 
Our results imply that, when Taurus expels its residual gas 
and becomes a pure $N$-body system, it will probably not 
simply collapse into a small, dense open cluster. Indeed, it 
is more likely that it will separate into two or three
small clusters, which in a few Myrs may look as though they 
formed  separately from each other. As we have shown, this 
depends crucially on the virial ratio of the final (gas free) 
cluster, and on the coherence of the velocity dispersion. 
Even a cluster which is initally fairly smooth in appearance, 
such as IC348 (Najita et al. 2000), may develop 
substructure and look very different in a few Myrs.

Second, the rapid dynamical evolution early in a cluster's 
life is also important when setting the initial conditions 
for $N$-body simulations of cluster evolution. Simple
initial conditions (such as Plummer spheres) will fail to 
capture this potentially important stage of cluster evolution. 
Processes like binary capture or dissolution, and stellar 
ejection, which probably occur in the first few Myr of a
cluster's lifetime, may be strongly affected by the evolving 
density substructure, and the dynamical changes we have 
identified here.


\begin{acknowledgements}
The {\sc grape-5} was purchased on PPARC grant PPA/G/S/1998/00642, SPG
is supported by PPARC grant PPA/G/S/1998/00623.
\end{acknowledgements}

\newpage

\newpage

\newpage

\newpage

\newpage

\newpage

\newpage

\newpage

\newpage

\end{document}